# The Realization of Virtual Environments in the Lower Limb Exoskeletal Robot

Minsu Chang, Doyoung Jeon

*Abstract*— This study proposes the realization of various virtual environments using a lower limb exoskeletal robot for futuristic gait rehabilitation. The proposed method allows the user to feel virtual gravity, buoyancy, and drag while actively walking. The virtual environments include four fluidic conditions of Water, Olive oil, Honey, and Peanut Butter, and four gravitational conditions consisting of the Earth's, Moon's, Mars', and Jupiter's gravity. The control method of the lower limb exoskeletal robot is as follows. First, torque feedback is applied to control the interaction force between the exoskeletal robot and its user. Second, the reference torque is computed in real time with the dynamic equations of the human body and the kinematic data. The eight environments were implemented via the EXOWheel, which is a wheelchair-integrated lower limb exoskeletal robot. While attaching the electromyography sensors and wearing the EXOWheel, eight healthy subjects walked actively under the virtual conditions. Experimental results show that the muscular force signals adequately change depending on the gravitational, buoyant, and drag effects. Blind tests were performed to predict the type of virtual environment, and the results showed that healthy people could distinguish eight virtual environments.

## I. Introduction

Exoskeletal robots are widely used for walking assistance and gait rehabilitation by people with difficulties such as stroke, Parkinson's disease, and spinal cord injury [1-6]. The functions of the rehabilitative robot consist of passive mode and active mode. In the passive mode, the robot repeatedly generates the patterns of motion without the user's contribution. In the case of the lower limb exoskeletal robot, various gait patterns can be realized by changing the robot's parameters such as step size, cadence, and joint's range of motion.

Clinical studies have shown better therapeutic results when the patients actively move, compared to receiving 100% of the support from the robot [7, 8]. Interesting and motivating methods have been developed so that patients can actively participate in robotic rehabilitation training. For example, Lokomat has game programs using avatars that mimic the user's motion [9]. ALEX displays the user's ankle position and targeted position of the ankle so that the user can actively follow the given gait pattern [10]. To increase the interest and concentration of the patient, the exoskeletal robot is combined with entertainment devices such as virtual reality (VR) or augmented reality [11].

People with weakened muscle strength or limited range of motion are trying to perform normal walking without physical support. The Assist-As-Needed paradigm (AAN) was proposed in which the exoskeletal robot can deliver adequate assistance and resistance, as the therapist's physical interaction. AAN is the active mode of the exoskeletal robot, and clinical studies have shown that AAN is adequate for the rehabilitation of abnormal gait pattern [12-14]. When the patient performs active motion along the given gait trajectory, AAN assists in case of insufficient force and restricts unnecessary movement. As an example of AAN, the force field controller implements the virtual damper and spring based on ankle position in the given gait pattern [15]. AAN delivers resistance or assistive force when the ankle position of the robot user moves away from the given trajectory.

Without the given trajectory for the robot-assisted gait training, a control algorithm that provides the interactive force appropriately has been developed. The control method achieves the effect of aquatic therapy by realizing the virtual fluidic environment. The walking assistive robot SUBAR generates the joint torque corresponding to the buoyancy and drag effect of water based on the dynamic equation and kinematic data of the lower limb [16-17]. Virtual fluidic environments with various levels were implemented in the previous study [18]. The lower limb exoskeletal robot Robogait realizes fluidic effects, and experimental results with healthy subjects showed that walking patterns varied with different levels of buoyancy and drag effect.

This study goes further than the previous studies and realizes eight virtual environments consisting of various gravity, buoyancy, and drag levels. The virtual environment is implemented via the EXOWheel, which has the functions of body weight support and overground walking. Previous studies have not shown results on whether the robot users feel the environmental effect. However, this study analyzes the muscular force signals measured by EMG sensors to verify that the gravity, buoyancy, and drag effect of the virtual environment are adequately implemented. Besides, the blind test verifies that the robot users can recognize the eight different virtual environments.

The remainder of the paper is structured as follows. The "Materials and Methods" section describes the torque feedback controller of the lower limb exoskeletal robot, EXOWheel, and the computation method for the joint torque command. The

Minsu Chang is with the Department of Mechanical Engineering, Sogang University, Seoul, Korea (e-mail: goldcal11@sogang.ac.kr).

Doyoung Jeon is with the Department of Mechanical Engineering, Sogang University, Seoul, Korea (correspondence e-mail: dyjeon@sogang.ac.kr).

"Results" section describes the simulation and experimental results for the eight types of environmental conditions. The "Conclusion" section concludes this study with the summarization of results and future works.

## II. Materials and Methods

In Fig. 1, the EXOWheel is designed for exercising and rehabilitative training in the activities of daily living of individuals with disabilities [19-20]. The user can utilize this device for transportation (i.e., an electric wheelchair) in the driving mode. As shown in the Fig. 1(a), the electric actuators located around the hip, knee, and ankle joints can provide the active/passive mode. The motion of this device is confined to the sagittal plane, and the other directional motion is possible with limited passivity. The body weight support system consists of an actuator and load cell so that the EXOWheel controls the lifting mechanism by measuring the user's weight. As shown in the Fig. 1(b), the user's active muscular torque can be estimated by the load cell integrated inside the exoskeletal robot [21]. The smart shoe with an insole pressure sensor located on the footplate can measure the vertical ground reaction force (GRF). During overground walking, the user's muscular torque can be estimated by removing the vertical GRF-induced torque from the joint torque measured by the load cell [22].

The following are discussed to enable the lower limb exoskeletal robot to implement the dynamic effects of the virtual environments. For simplicity, the following assumptions are made:
1) A human leg consists of two rigid cylinders (i.e., thigh and lower leg), and each segment is connected to a fixed hinge joint (i.e., hip joint and knee joint).
2) The model considers motion in the sagittal plane only.
3) The shape of each segment is assumed to be cylindrical, and its mass is uniformly distributed.
4) The human leg is fully immersed in the fluid. The fluid does not move, and the relative flow velocity is generated by the leg motion only.
5) The effects of vertical GRFs on hip and knee joint torque are compensated through the insole pressure sensors.

With the previously stated assumptions, the Fig. 2(a). shows the two-dimensional lower limb model in this study. In the figure, the subscripts 1 and 2 denote the hip and knee joints, respectively. In the real world, the equation of motion for the human lower limb model is expressed as

$$\boldsymbol{M}_\mathrm{H}(\boldsymbol{\theta})\ddot{\boldsymbol{\theta}} + \boldsymbol{V}_\mathrm{H}(\boldsymbol{\theta},\dot{\boldsymbol{\theta}}) + \boldsymbol{G}_\mathrm{H}(\boldsymbol{\theta}) = \boldsymbol{\tau}_\mathrm{M} + \boldsymbol{\tau}_\mathrm{E} \qquad (1)$$

where $\theta, \dot{\theta}, \ddot{\theta} \in \mathbb{R}^2$ are the joint angle, angular velocity, and angular acceleration, respectively. The subscript H denotes the human subject. $\boldsymbol{M}(\boldsymbol{\theta}) \in \mathbb{R}^{2\times2}$, $\boldsymbol{V}(\boldsymbol{\theta},\dot{\boldsymbol{\theta}}) \in \mathbb{R}^2$, and $\boldsymbol{G}(\boldsymbol{\theta}) \in \mathbb{R}^2$ represent the symmetric positive definite inertial matrix, Coriolis force, and gravitational torques, respectively. $\tau_\mathrm{M} \in \mathbb{R}^2$ is the vector of muscular torques, $\tau_\mathrm{E} \in \mathbb{R}^2$ is the vector of external torques, $\theta = [\theta_1\ \theta_2]^\mathrm{T}$, and $\tau = [\tau_1\ \tau_2]^\mathrm{T}$. The equation of motion for the human lower limb in the virtual environment is expressed as follows:

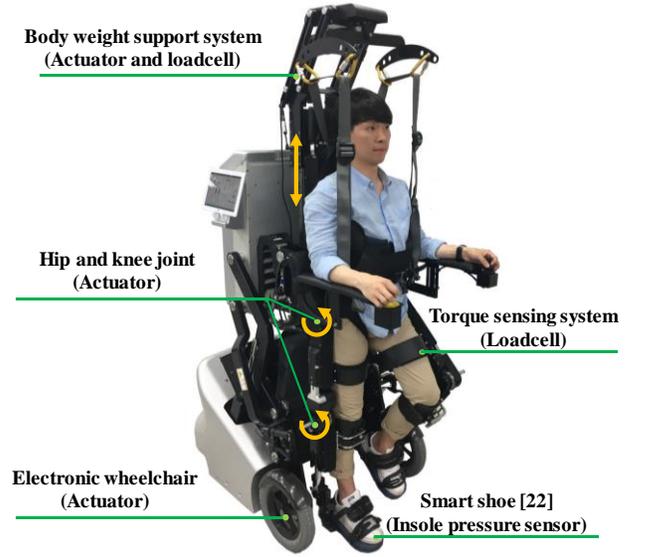

(a) The hardware specification of body weight support system, joint actuators, and sensor systems.

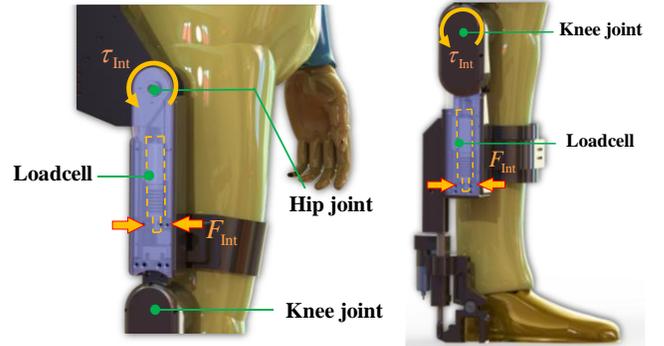

(b) Torque-sensing system of hip and knee joint to measure the interactive torque between the exoskeletal robot and user.

Fig. 1. Wheelchair integrated lower limb exoskeletal robot, EXOWheel.

$$\boldsymbol{M}_\mathrm{H}(\boldsymbol{\theta})\ddot{\boldsymbol{\theta}} + \boldsymbol{V}_\mathrm{H}(\boldsymbol{\theta},\dot{\boldsymbol{\theta}}) + \boldsymbol{G}^*(\boldsymbol{\theta}) + \boldsymbol{B}^*(\boldsymbol{\theta}) + \boldsymbol{D}^*(\boldsymbol{\theta},\dot{\boldsymbol{\theta}}) = \boldsymbol{\tau}_\mathrm{M} \qquad (2)$$

where the star (*) represents the virtual environments realized by the exoskeletal robot. $\boldsymbol{B}(\boldsymbol{\theta}) \in \mathbb{R}^2$ and $\boldsymbol{D}(\boldsymbol{\theta},\dot{\boldsymbol{\theta}}) \in \mathbb{R}^2$ are the vectors of buoyant torques and drag torques, respectively.

This study aims to implement the virtual environment as given by Eq. (2) by properly organizing the external torque ($\tau_\mathrm{E}$) of Eq. (1). When a person wears the exoskeletal robot, the $\tau_\mathrm{E}$ is the disturbance produced by the robot and the GRF. The $\tau_\mathrm{E}$ is expressed as follows:

$$\boldsymbol{\tau}_\mathrm{E} = \boldsymbol{\tau}_\mathrm{R} - \left(\boldsymbol{M}_\mathrm{R}(\boldsymbol{\theta})\ddot{\boldsymbol{\theta}} + \boldsymbol{V}_\mathrm{R}(\boldsymbol{\theta},\dot{\boldsymbol{\theta}}) + \boldsymbol{G}_\mathrm{R}(\boldsymbol{\theta}) + \boldsymbol{\tau}_\mathrm{GRF}\right) \qquad (3)$$

where the subscript R denotes the exoskeletal robot, $\tau_\mathrm{R} \in \mathbb{R}^2$ is the vector of the torques generated by the robot's actuators, and $\tau_\mathrm{GRF} \in \mathbb{R}^2$ is the vector of the vertical GRF-induced torque. If Eq. (3) is substituted in Eq. (1), it is expressed as

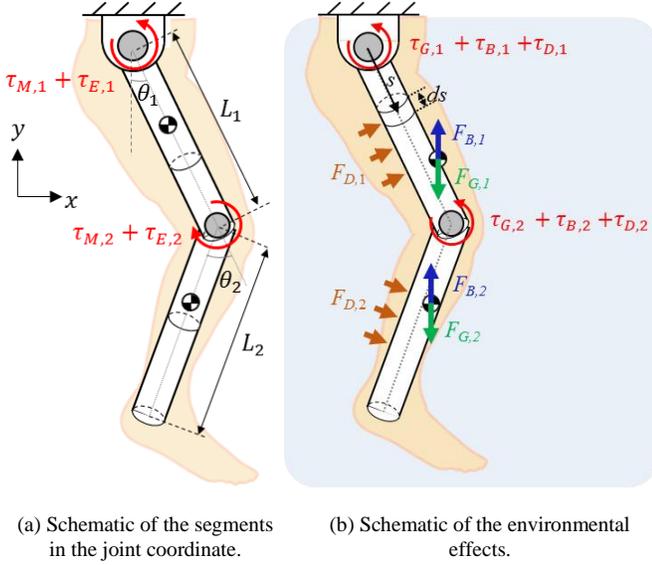

(a) Schematic of the segments in the joint coordinate.   (b) Schematic of the environmental effects.

Fig. 2. Two-dimensional model of the human lower limb used in this study.

$$\begin{aligned}\boldsymbol{\tau}_M =& \left(\boldsymbol{M}_H(\boldsymbol{\theta})\ddot{\boldsymbol{\theta}} + \boldsymbol{V}_H(\boldsymbol{\theta},\dot{\boldsymbol{\theta}}) + \boldsymbol{G}_H(\boldsymbol{\theta})\right) \\ &+ \left(\boldsymbol{M}_R(\boldsymbol{\theta})\ddot{\boldsymbol{\theta}} + \boldsymbol{V}_R(\boldsymbol{\theta},\dot{\boldsymbol{\theta}}) + \boldsymbol{G}_R(\boldsymbol{\theta}) + \boldsymbol{\tau}_{GRF}\right) \\ &- \boldsymbol{\tau}_R \end{aligned} \quad (4)$$

As seen from the Eq. (4), the robot user is affected by a disturbance such as the vertical GRF and the exoskeleton's inertia, Coriolis, and gravity. Thus, the robot's actuator should generate the compensating torque for $\boldsymbol{M}_R(\boldsymbol{\theta})$, $\boldsymbol{V}_R(\boldsymbol{\theta},\dot{\boldsymbol{\theta}})$, $\boldsymbol{G}_R(\boldsymbol{\theta})$, and $\boldsymbol{\tau}_{GRF}$. The actuating torque for compensation is expressed as

$$\boldsymbol{\tau}_{Comp} = \hat{\boldsymbol{\tau}}_{GRF} + \hat{\boldsymbol{M}}_R(\boldsymbol{\theta})\ddot{\boldsymbol{\theta}} + \hat{\boldsymbol{V}}_R(\boldsymbol{\theta},\dot{\boldsymbol{\theta}}) + \hat{\boldsymbol{G}}_R(\boldsymbol{\theta}) \quad (5)$$

where the hat (^) means the estimated value. To realize the virtual environment represented by the Eq. (2), the inertia and Coriolis of the user's lower limb are maintained, and the leg's gravity, buoyancy, and drag should be changed as desired. In summary, the joint torque generated by the robot's actuator is expressed as

$$\begin{aligned}\boldsymbol{\tau}_R =& \left(\hat{\boldsymbol{G}}_H(\boldsymbol{\theta}) + \hat{\boldsymbol{B}}_H(\boldsymbol{\theta}) + \hat{\boldsymbol{D}}_H(\boldsymbol{\theta},\dot{\boldsymbol{\theta}})\right) + \boldsymbol{\tau}_{Comp} \\ &- \left(\boldsymbol{G}^*(\boldsymbol{\theta}) + \boldsymbol{B}^*(\boldsymbol{\theta}) + \boldsymbol{D}^*(\boldsymbol{\theta},\dot{\boldsymbol{\theta}})\right)\end{aligned} \quad (6)$$

By substituting the Eq. (6) in the Eq. (4), the governing equation becomes the Eq. (2), which means that the robot user can feel the virtual environment.

Fig. 3 represents the block diagram of the torque feedback control to implement the environmental effect using the exoskeletal robot. The actuator applies the $\tau_{comp}$ to compensate for the static and coulomb frictions and the weight of the exoskeleton. For the details of the gravity and friction compensation, please refer to [23]. The reference torque ($\tau_{ref}$) is the summation of the desired torque for the impedance control ($\tau_{imp}$) and the environmental torque corresponding to the gravity, buoyancy, and drag of the virtual environment.

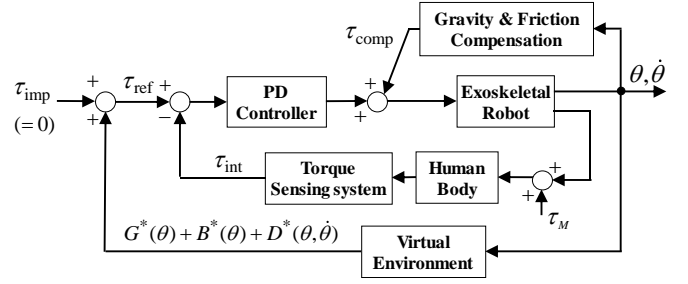

Fig. 3. Block diagram of the overall control structure.

Note that zero impedance should be implemented on the robot's joint for the user's free motion, so the $\tau_{imp}$ becomes zero. The interaction torque is measured by the force-sensing system in the Fig. 1(b). The PD controller of the torque control loop is designed to minimize the error between the $\tau_{int}$ and the $\tau_{ref}$.

In the Eqs. (1–6), **M**, **V**, and **G** are characterized by the segment inertial parameters and can be written as

$$\boldsymbol{M}(\boldsymbol{\theta}) = \begin{bmatrix} M_{11}(\boldsymbol{\theta}) & M_{12}(\boldsymbol{\theta}) \\ M_{21}(\boldsymbol{\theta}) & M_{22}(\boldsymbol{\theta}) \end{bmatrix} \quad (7)$$

$$\boldsymbol{V}(\boldsymbol{\theta},\dot{\boldsymbol{\theta}}) = \begin{bmatrix} -L_1(X_2 \sin\theta_2 + Y_2 \cos\theta_2)\cdot(\dot{\theta}_2^2 + 2\dot{\theta}_1\dot{\theta}_2) \\ L_1(X_2 \sin\theta_2 + Y_2 \cos\theta_2)\cdot\dot{\theta}_1^2 \end{bmatrix} \quad (8)$$

$$\boldsymbol{G}(\boldsymbol{\theta}) = g\begin{bmatrix} X_1 \sin\theta_1 + Y_1 \cos\theta_1 + X_2 \sin(\theta_{12}) + Y_2 \cos(\theta_{12}) \\ X_2 \sin(\theta_{12}) + Y_2 \cos(\theta_{12}) \end{bmatrix} \quad (9)$$

where the subscript 12 denotes the summation of hip joint angle and knee joint angle (i.e. $\theta_{12} = \theta_1 + \theta_2$), and $g$ is the gravitational acceleration as shown in Fig. 2(a). The segment inertial parameters ($J_1$, $J_2$, $X_1$, $X_2$, $Y_1$, and $Y_2$) of the human subject are obtained through the experiments using the EXOWheel and the experimental process is provided in [20, 21].

As shown in the Fig. 2(b), the buoyant force exerted on the $i$-th segment, $F_{b,i}$, can be calculated using Archimedes' principle as

$$\boldsymbol{F}_{B,i} = \rho_F g V_i \hat{t} = \frac{\rho_F}{\rho_H} m_i g \hat{t} \quad (10)$$

where $\rho_F$ is the mass density of the fluid, $\rho_H$ is the mass density of a human body (approximately 1041 kg/m$^3$ [24]), $V_i$ is the volume of the segment, and $\hat{t}$ is the unit normal vector pointing outward from the ground. Note that the buoyant torque works opposite to the gravitational torque, and its amplitude can be calculated as the scaled weight to the gravitational torque. From the Eq. (10), the joint torque induced by the buoyant force can be computed as

$$\boldsymbol{B}(\boldsymbol{\theta}) = -\frac{\rho_F}{\rho_H}\boldsymbol{G}(\boldsymbol{\theta}) \quad (11)$$

In fluid dynamics, the quadratic drag equation is commonly used to determine the drag force on an object when the drag coefficient is known [25], and the equation is

$$F_D = -\frac{1}{2} C_D \rho_F v^2 A \tag{12}$$

where $v$ is the flow velocity, $F_D$ is the drag force in the direction of the flow velocity, and $A$ is the reference area. The drag coefficient, $C_D$, is a function of the Reynolds number, Re, for a certain body shape, and the computational model for a smooth cylinder [25] is expressed as

$$C_D = 1.18 + \frac{6.8}{\text{Re}^{0.89}} + \frac{1.96}{\text{Re}^{0.5}} - \frac{0.0004\,\text{Re}}{1 + 3.64 \times 10^{-7}\,\text{Re}^2} \tag{13}$$

$$\text{Re} = \frac{\rho_F l}{\mu_F} v \tag{14}$$

where $l$ is the characteristic length of the rigid cylinder, $\rho_F$ is the mass density of the fluid, $v$ is the flow velocity, and $\mu_F$ is the dynamic viscosity of the fluid.

By using the Eq. (12), the drag force exerted on a small element of the $i$-th segment, $d\mathbf{F}_{d,i}$, is obtained as

$$d\mathbf{F}_{D,i} = -\rho_F C_D r_i \mathbf{v}_i(s) |\mathbf{v}_i(s)| ds_i \tag{15}$$

where $s_i$ is the position vector from a joint $i$ to the element, $r_i$ is the radius of the cross-section at $s$, and the reference area, $A$, is equal to $2 r_i s_i$. The joint torques, which are induced by the drag forces at the small element, are

$$d\boldsymbol{\tau}_{D,1} = \mathbf{s}_1 \times d\mathbf{F}_{D,1} = \hat{\mathbf{z}}_1 \left[ -\rho_F C_D r_1 s_1 (s_1 \dot{\theta}_1)^2 \operatorname{sgn}(\dot{\theta}_1) \right] ds_1 \tag{16}$$

$$d\boldsymbol{\tau}_{D,2} = \hat{\mathbf{z}}_2 \left[ -\rho_F C_D r_2 s_2 \begin{pmatrix} L_1^2 \dot{\theta}_1^2 \operatorname{sgn}(\dot{\theta}_1) + s_2^2 \dot{\theta}_{12}^2 \operatorname{sgn}(\dot{\theta}_{12}) \\ + L_1 s_2 \dot{\theta}_1 \dot{\theta}_{12} \cos\theta_2 (\operatorname{sgn}(\dot{\theta}_1) + \operatorname{sgn}(\dot{\theta}_{12})) \end{pmatrix} \right] ds_2 \tag{17}$$

where $\hat{\mathbf{z}}_i$ is a unit vector normal to the sagittal plane. Here, the dot product of the two vectors $\mathbf{s}_i$ and $d\mathbf{F}_{d,i}$ yields the sign function $\operatorname{sgn}(\dot{\theta})$. Integrating the Eqs. (16)–(17) over the length of the thigh and lower leg, respectively, the total joint torques due to the drag force are formulated as the Eq. (18).

In this research, four different fluidic conditions (water, olive oil, honey, and peanut butter) and four different gravitational conditions (gravities of the Earth, Moon, Mars, and Jupiter) are selected as the virtual environments. To obtain the gravitational, buoyant, and drag torque of the virtual environments, the properties of the eight environments are listed in Table I. The density and the dynamic viscosity of the fluids are determined with the following assumptions: the temperature of the environment is 20 °C; the flow velocity, $v$, is 2 km/h, which is the general walking speed of a healthy person; and the characteristic length, $l$, is 0.2 m, which is the diameter of a healthy male's thigh [24]. The Reynolds number (Re) and drag coefficient ($C_D$) are determined with the dynamic properties in Table I and Eqs. (13)–(14).

TABLE I
PROPERTIES OF THE VIRTUAL ENVIRONMENTS IN THIS RESEARCH.

| Environment | $g^{(1)}$ (m/s²) | $\rho_F^{(2)}$ (kg/m³) | $\mu^{(3)}$ (Pa·s) | Re$^{(4)}$ | $C_D^{(5)}$ |
|---|---|---|---|---|---|
| Water | 9.81 | 998.2 | 1.00E-03 | 109800 | 1.17 |
| Olive oil |  | 800 | 1.00E-01 | 880 | 0.99 |
| Honey |  | 1420 | 5 | 31 | 1.84 |
| Peanut butter |  | 1283 | 250 | 0.56 | 15.1 |
| The Earth | 9.81 | 1.20 | 1.80E-04 | 733 | 1.03 |
| Moon | 1.63 |  |  |  |  |
| Mars | 3.71 |  |  |  |  |
| Jupiter | 24.5 |  |  |  |  |

$^{(1)}$ Gravitational acceleration. $^{(2)}$ The density of fluid (The mass density of a human body is approximately 1041 kg/m³ [24]). $^{(3)}$ Dynamic viscosity of the fluid. $^{(4)}$ Reynolds numbers from Eq. (14). $^{(5)}$ The drag coefficient from Eq. (13).

Fig. 4 shows the simulation results for joint angle, angular velocity, and torque consisting of the gravity, buoyancy, and drag. In the simulation, the joint torques are computed for the normal walking [26] at 2 km/h in eight different virtual environments. Fig. 4(a) shows the kinematic data of the hip and knee joint during one gait cycle (GC) of normal walking. The data of the joint angle and joint angular velocity referred to are from the OpenSim program. The OpenSim is a widely used software for anthropometric simulation, including the motion of the human body, GRF, muscle strength, and joint torques [27]. Figs. 4(b)–(e) show the hip and knee joint torque computed for the eight different virtual environments based on the Eqs. (9)–(18) and Table I. In the Figs. 4(b)–(d), the buoyant and drag torques are close to zero in the gravitational environments of the Earth, Moon, Mars, and Jupiter, while they increase significantly depending on fluidic atmospheres such as Water, Olive oil, Honey, and Peanut butter. Because the velocity of leg motion is faster in the swing phase (Gait Cycle 60–100%) than in the stance phase, as shown in Fig. 4(a), the drag torque in Fig. 4(d)–(e). is dominant in the swing phase.

$$\boldsymbol{\tau}_{D,1} = \int_0^{L_1} d\boldsymbol{\tau}_{D,1} = -\rho_F C_D r_1 \left( \frac{L_1^4 \dot{\theta}_1^2}{4} \right) \operatorname{sgn}(\dot{\theta}_1)$$

$$\boldsymbol{\tau}_{D,2} = \int_0^{L_2} d\boldsymbol{\tau}_{D,2} = -\rho_F C_D r_2 \left( \frac{L_1^2 L_2^2 \dot{\theta}_1^2}{2} \operatorname{sgn}(\dot{\theta}_1) + \frac{L_2^4 \dot{\theta}_{12}^2}{4} \operatorname{sgn}(\dot{\theta}_{12}) + \frac{L_1 L_2^3 \dot{\theta}_1 \dot{\theta}_{12} \cos\theta_2}{3} (\operatorname{sgn}(\dot{\theta}_1) + \operatorname{sgn}(\dot{\theta}_{12})) \right) \tag{18}$$

$$\mathbf{D}(\boldsymbol{\theta}, \dot{\boldsymbol{\theta}}) = \begin{bmatrix} \tau_{D,1} - \tau_{D,2} \\ \tau_{D,2} \end{bmatrix}$$

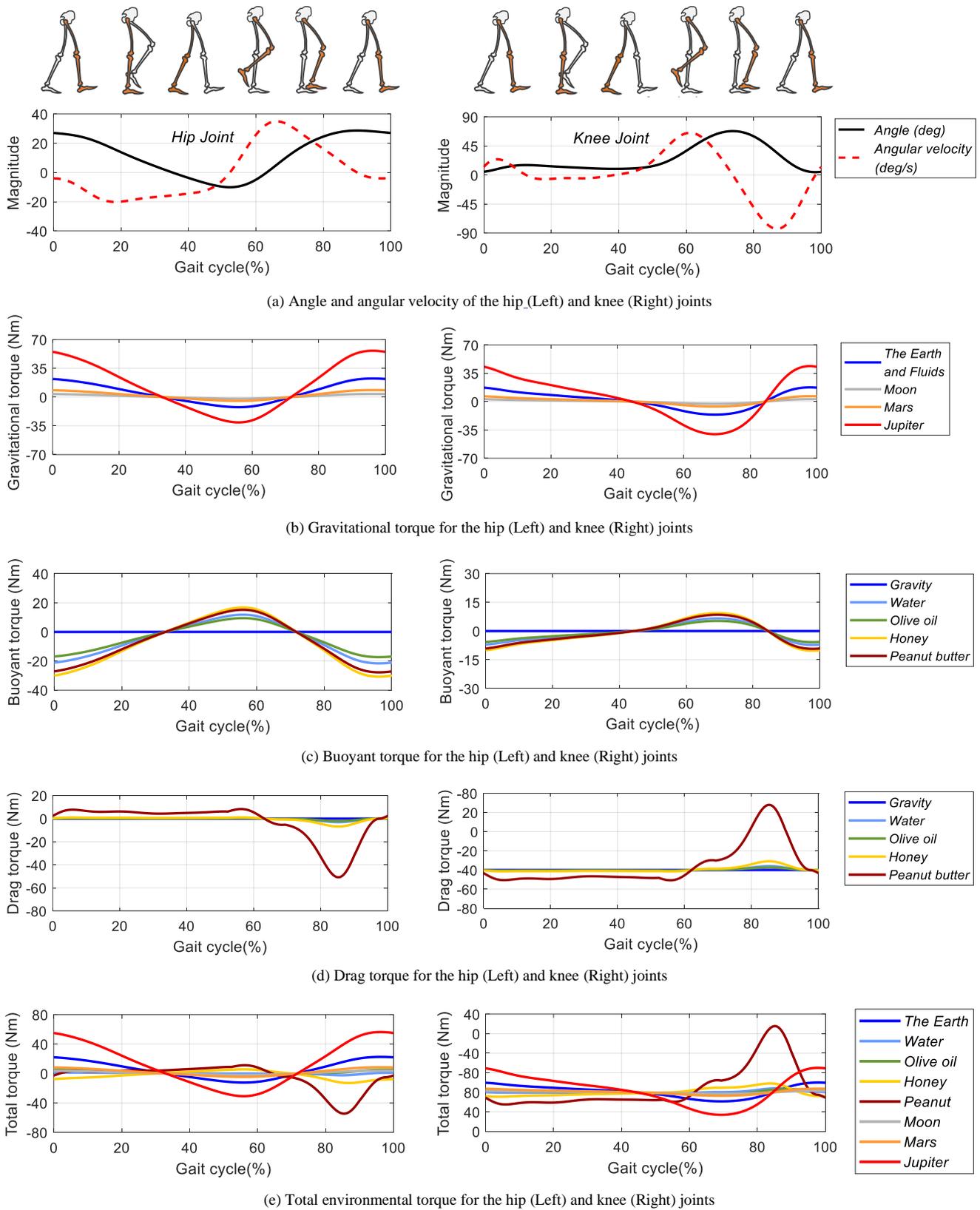

(a) Angle and angular velocity of the hip (Left) and knee (Right) joints

(b) Gravitational torque for the hip (Left) and knee (Right) joints

(c) Buoyant torque for the hip (Left) and knee (Right) joints

(d) Drag torque for the hip (Left) and knee (Right) joints

(e) Total environmental torque for the hip (Left) and knee (Right) joints

Fig. 4. Simulation results of joint torque corresponding to the eight different virtual environments in a gait cycle.

To analyze the quantitative variation of the gravitational, buoyant, and drag torques according to the different virtual environments, the root-mean-square (RMS) value of each component is computed from the simulation results. The results are presented in Table II. Under the gravities of the Earth, Moon, Mars, and Jupiter, the gravitational torques are dominant, while the buoyant and drag torques are close to zero. In Water, Olive oil, Honey, and Peanut butter, the buoyant torque is close to or more prominent than the gravitational torque. These two torques are canceled because they act on the user's joint in the opposite direction. The dominant drag torques are observed in fluids that have large drag coefficients, such as honey and peanut butter. In the case of peanut butter, for example, the RMS value of the drag torque is twice that of the gravitational torque; therefore, the user finds it difficult to move their leg. As part of the simulation results, environmental torques under the eight virtual conditions during walking with a speed of 2 km/h are adequately calculated.

To verify the virtual environments realized by the EXOWheel, eight healthy subjects (Mean ± SD; weight = 70.2 ± 5.6 kg, height = 174.6 ± 3.4 cm, age = 25.7 ± 2.7 years) wearing the EXOWheel performed overground walking with the speed of 2 km/h. Experiments were approved by the Ethics Committee of Sogang University and written informed consent was obtained from all participants. A monitor connected to the EXOWheel displays the desired and actual position of the subject's ankle, since the visual feedback is effective in enhancing and motivating the robotic gait training [29–31].

This study investigates a subject's muscle activation according to the virtual environments. The electromyography (EMG) data from the surface electrodes were recorded for two biarticular muscles of the left leg: rectus femoris (RF) for hip flexion and knee flexion, and hamstrings (HAM) for hip extension and knee flexion. The two muscles are biarticular muscles that cross the hip and knee joints at the front (RF) and back (HAM) of the thigh, and the muscles are essential for many lower limb activities. The surface electrodes were attached according to the SENIAM guidelines [32]. The raw EMG signals from the electrodes were amplified with the Bagnoli$^{TM}$ 8-channel system (Delsys Inc.) with a gain of 1,000 and bandpass filtered (Butterworth, fourth-order, 30–450 Hz), full-wave rectified, and then low-pass filtered (Butterworth, second-order, 4 Hz) to obtain the linear EMG envelopes. To calculate the muscular torque from the EMG signals, the model reported by Olney and Winter [33] was used. For the details of the isometric calibration procedures, see [21–22, 34–36].

### III. EXPERIMENTAL RESULTS AND DISCUSSION

For the comparison of the experimental results with the simulation results, this study applies the estimation method for the muscle force of the lower limb during walking. From the joint torque of the hip and knee, the strength of the seven muscle groups of the lower limb can be estimated [32]. In this study, the muscular forces of RF and HAM muscle are estimated from the environmental torque (Fig. 4(e)), and they are compared with the EMG signals.

Fig. 5 and Fig. 6 show the comparison between the simulation results and experimental results. In these figures, the black line is the muscle force from the simulation in the Fig. 4, the red line represents the muscle force from the mean value of the EMG envelope, and the red error bar is the standard deviation. The similarity between the measured EMG envelope and the estimated muscle force is evaluated by the Pearson correlation coefficient $r$, and the results indicate that the realization of virtual environments can induce muscle activation patterns as the simulation. The comparison between the estimated muscle forces and measured EMG envelope verifies the fidelity of the virtual environments realized by the EXOWheel.

TABLE II
ROOT MEAN SQUARE (RMS) TORQUE FOR THE VIRTUAL ENVIRONMENTS FROM THE SIMULATION RESULTS (UNIT: NM).

| Joint | Environment | Gravity | Buoyancy | Drag |
|---|---|---|---|---|
| Hip | Water | 12.88 | 12.35 | 2.03 |
|  | Olive oil |  | 9.90 | 1.37 |
|  | Honey |  | 17.56 | 4.5 |
|  | Peanut butter |  | 15.87 | 33.47 |
|  | The Earth's gravity |  | 0.01 | 0 |
|  | Moon's gravity | 2.14 |  |  |
|  | Mars' gravity | 4.87 |  |  |
|  | Jupiter's gravity | 32.16 |  |  |
| Knee | Water | 4.56 | 4.37 | 2.06 |
|  | Olive oil |  | 3.50 | 1.39 |
|  | Honey |  | 6.22 | 4.57 |
|  | Peanut butter |  | 5.62 | 33.99 |
|  | The Earth's gravity |  | 0.01 | 0 |
|  | Moon's gravity | 0.76 |  |  |
|  | Mars' gravity | 1.72 |  |  |
|  | Jupiter's gravity | 11.38 |  |  |

In 40–80% gait cycle of the Fig. 5(a), the HAM muscle generated the muscular force for the knee flexor to overcome the Earth's gravity, and a high correlation between the simulation and experiment was obtained ($r$=0.91). Figs. 5(b)–(d) show the increasing peak value for the HAM muscle force as the gravity of the virtual environment grows. According to Table II, the buoyant torques of water and olive oil are weaker than the Earth's gravity. Thus, HAM muscle forces in the Figs. 5(e)–(f) (Water and Olive oil) tend to be related to the forces in the Figs. 5(b)–(c) (Moon and Mars). Under Honey and Peanut butter in the Figs. 5(g)–(h), a high magnitude of HAM muscle strength is required for the hip extensor due to the strong buoyancy. Thus, the strong HAM muscle strength is generated at 0–40% and 80–100% gait cycle where the hip extensor occurs. Under Peanut butter in Fig. 5(h), the strong drag torque results in the high magnitude of the HAM muscle force. RF muscle force in Fig. 6 also shows a trend similar to that of the HAM muscle strength in each virtual environment.

To verify whether the subjects can recognize the types of virtual environments, the users were blinded to the experimental condition and chose one candidate that they felt closest among the eight environments. Table III gives the success rate of distinguishing the environments through blind tests.

In most cases, the subjects can feel the difference between the experimental conditions. The subjects can distinguish the environments correctly when the generated torque is significantly different from the others. Besides, the blind tests confirm that people can distinguish and recognize the virtual environment from the exoskeletal robot only by the sense of touch, but water, olive oil, honey, Moon, and Mars have a success rate of less than 100 %.

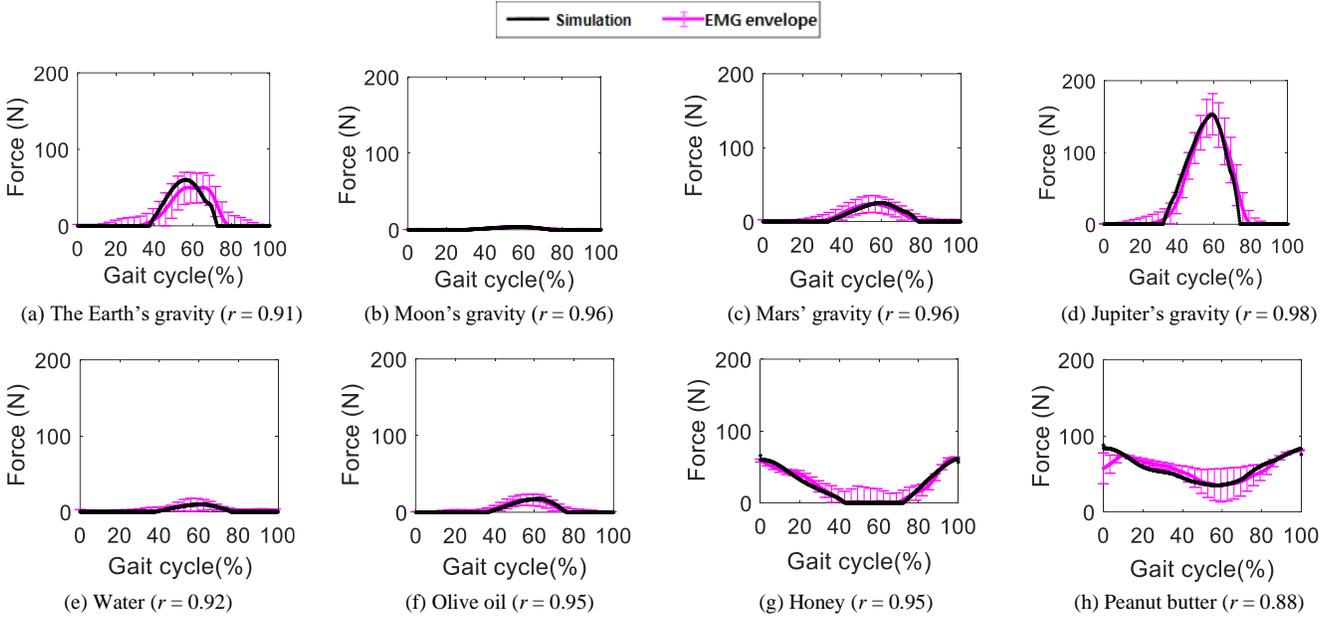

Fig. 5. Muscle forces of HAM from the simulation and experiments under the eight different virtual environments.

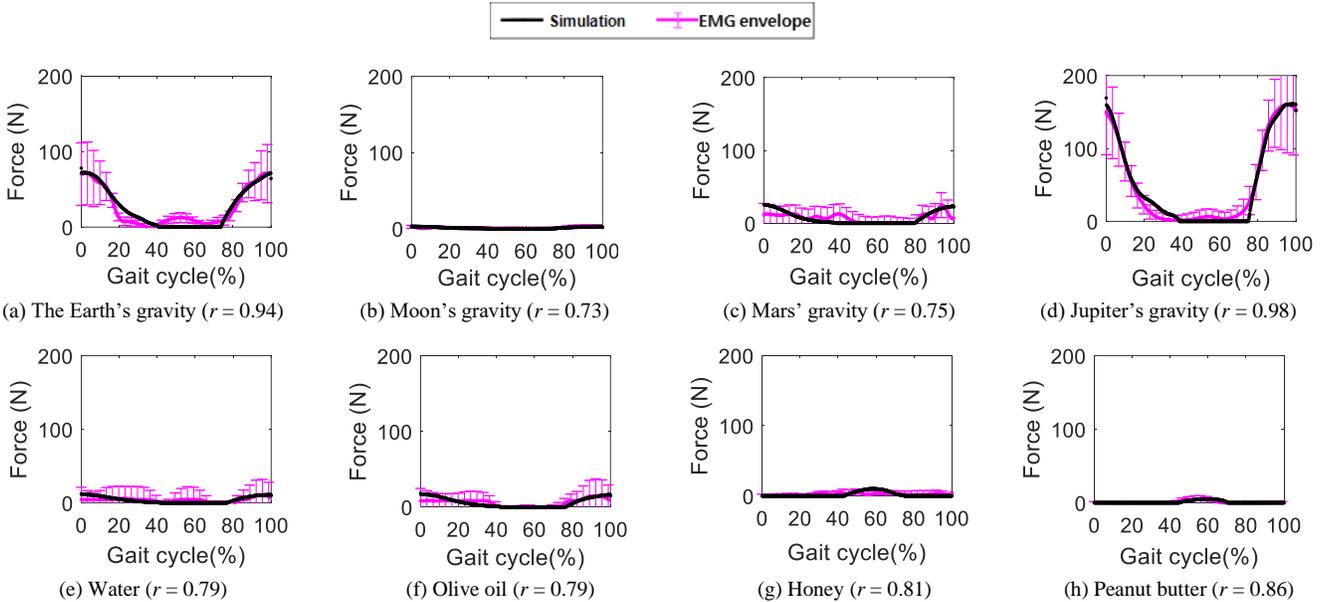

Fig. 6. Muscle forces of RF from the simulation and experiments under the eight different virtual environments.

The reasons why the results of the blind tests are shown can be explained as follows. First, the buoyant torque and drag torque are similar in the case of water and olive oil, so they cannot be distinguished only by the joint torque delivered from the robot. Second, as the gravity of the Moon and Mars are approximately 1/6 and 1/3 of the Earth's gravity, respectively, it is difficult to tell the difference between the two types of gravity that are lighter than the Earth's unless the loads are added to the lower limb. However, in a real environment, many important factors, such as temperature, pressure, and skin, affect the actual "feeling" of the environment.

## IV. CONCLUSION

This study implements the environmental effects using a lower limb exoskeletal robot so that users can feel the virtual environment. The virtual environments are implemented by the wheelchair-integrated lower limb exoskeletal robot, EXOWheel. The EXOWheel computes gravitational, buoyant, and drag torques in real time based on the dynamic model of the lower limb and kinematic data, and the environmental torques are applied to the user by torque feedback control. The experimental results using EMG sensors and blind tests have shown that subjects wearing the EXOWheel can feel and recognize the eight types of virtual environments.

The virtual environment is expected to apply not only to personal entertainment but also to more intensive lower limb training during robot-assisted therapy for rehabilitation. In a future study, the virtual environment will incorporate a 3D VR system to enhance the interest in robotic gait training, and also create a more realistic environment by adding some ways to spray smells or fluids.